# ESTIMATING TRANSIT VEHICLE DWELL TIMES AT BUS STOPS

**Taqwa AlHadidi**
Charles E. Via, Jr. Department of Civil and Environmental Engineering
Center for Sustainable Mobility, Virginia Tech Transportation Institute
3500 Transportation Research Plaza
Blacksburg, VA 24061
Tel: (540) 231-1058; Fax: (540) 231-1555; Email: Taqwaa1@vt.edu
ORCID number: 0000-0001-6388-0904

**Hesham A. Rakha (Corresponding author)**
Charles E. Via, Jr. Department of Civil and Environmental Engineering
Center for Sustainable Mobility, Virginia Tech Transportation Institute
3500 Transportation Research Plaza
Blacksburg, VA 24061
Tel: (540) 231-1505; Fax: (540) 231-1555; Email: hrakha@vt.edu
ORCID number: 0000-0002-5845-2929

**ABSTRACT**
This work presents a quantitative approach to estimate the total time spent in the vicinity of a bus stop including the deceleration time, the boarding and alighting time (developed in an earlier study), the acceleration time, and re-entry time (time required to merge into the adjacent lane). Different statistical models were used to compute the deceleration, acceleration and merge times. Typical deceleration and acceleration levels were computed using kinematic equations that were then used to compute both the deceleration and acceleration times. The adopted method to estimate both the deceleration time and the acceleration time was validated utilizing transit data from Blacksburg Transit (BT) using the mean absolute percentage error (MAPE) and root mean square error (RMSE). The MAPE and RMSE values were calculated to be 0.3% and 13.3% for the deceleration time and 0.42% and 2.72% for the acceleration time, respectively. The re-entry time was estimated to be a function of the adjacent roadway traffic density using both a multiple linear regression and Bayesian regression approach. Both methods showed consistency in estimating the merge-time model coefficients. The proposed models can be integrated with transit applications to estimate transit vehicle travel times.

## INTRODUCTION
Public transit buses, compared to private vehicles, provide an efficient transportation mode and help in reducing traffic congestion and energy/fuel consumption. However, the accurate estimation of travel time is still a challenging task due to different interacting factors that affect it. These factors include the interaction with other transportation modes (i.e., cars, rails and bicycles), the variability in transit vehicle stop time at bus stops, the stochasticity of travel time between two stops, and the stochasticity in demand patterns.

The accurate estimation of bus travel and arrival times using emerging technology including automated vehicle location (AVL), detectors, and probe vehicles can assist in providing real-time information about bus arrival times for transit system users, agencies, and operators. This, in turn, would help in attracting more riders, improving users' satisfaction, and reducing travel time with its associated cost.

In essence, estimation of the bus arrival time at any stop depends mainly on the time that a transit vehicle spends at a stop and the travel time between two stops. The travel time components between two stops have been quantified using data from different states in the 1980s, which are shown in Figure 1. The time at bus stops has been estimated using a conventional method (i.e., by considering only the boarding and alighting times and doors open-close time) (1). However, a transit vehicle may spend longer than this anticipated time such that dwell time does not become only a function of boarding/alighting time, but also has other components as shown in Figure 2 (2).

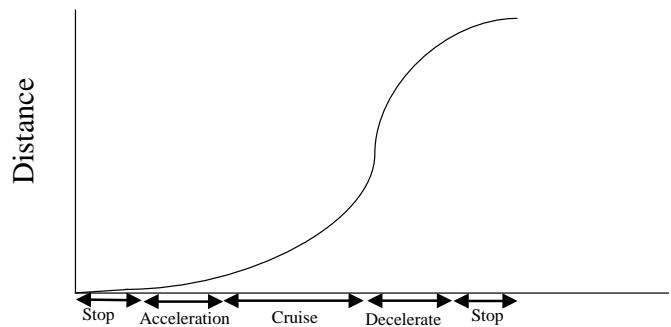

Figure 1: Transit travel time components between stops (1)

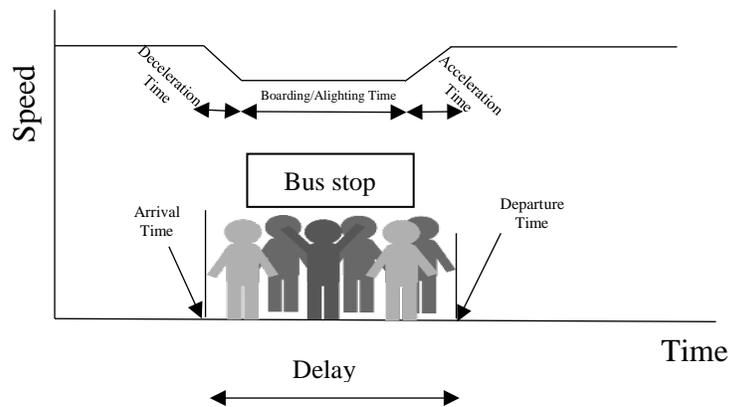

Figure 2: Transit travel time components at bus stop (2)

As can be seen in Figure 2, the total time spent at a bus stop includes the time it takes the transit vehicle to completely stop, the boarding/alighting time, the acceleration time, and the merge time. The last two components are known, according to the Transit Capacity and Quality of Service Manual (TCQSM) , as re-entry delay especially at the off-curb stops (2). The re-entry delay expresses an imposed delay on the transit vehicle, which is caused by either the queued traffic in front of the transit vehicle or the traffic volume in the adjacent lane that prevents the transit vehicle from merging easily with traffic. Adding this time to the boarding/alighting time that has been studied by the researchers in a previous study helps in accurate estimations of bus dwell times at bus stops.

Speaking of the acceleration/deceleration time, adding the transit vehicle deceleration/acceleration performance characteristics assists in providing an in-depth understanding of vehicle deceleration/acceleration behavior that could be used to provide a more accurate estimation of buses' arrival times at bus stops and signal stops as well update the current bus bay configuration requirements.

## STUDY OBJECTIVES AND PAPER LAYOUT

The contribution of this paper to the literature is three-fold. First, it models transit vehicle deceleration/acceleration behavior considering a kinematic model based on field data by explicitly modeling the driver and the vehicle characteristics. Second, it estimates the transit vehicle re-entry time, which is a vital component of estimating the dwell time, especially at off-curb bus stops. Third, it explicitly accounts for the all components of the bus dwell time at a bus stop. Results of this work can be incorporated within microscopic simulations to estimate the total time spends in the vicinity of bus stops and predict bus arrival and departure times at bus stops. The results of this research can be extended to update bus bay design configurations, signal timings, bus schedules, and bus arrival times.

The paper is laid out as follows. Initially, previous research findings are presented. Subsequently, the description of the data collection and reduction is described followed by a fully detailed description of the adopted methodology. Next, a discussion of modeling the acceleration and deceleration later that was used later to analytically compute the acceleration and deceleration times is presented, followed by a discussion for modeling the merge duration using a classical frequentist statistical model and the stochasticity in the model parameters using the Bayesian regression. Subsequently, the results of modeling the different delay components at a bus stop are presented. A discussion of the study results is presented next. Finally, the conclusions of the paper are drawn and insights for future work are provided.

## PREVIOUS FINDINGS

The aim of this literature review is to provide a summary of current models that model the transit lost time components at the bus stops considering transit deceleration and acceleration times, in addition to re-entry delay. As such, the lost time between bus stopping time and the beginning of passenger boarding/alighting is beyond the scope of this work.

### Literature review for bus dwell time

Several works modeled the dwell time in different methods and using several data. Yet, none of them modeled the different dwell time components. Ironically, all of these models estimate the dwell time as function of boarding/alighting time(BA). However, a study was conducted previously by the researchers (3) to estimate the BA times as a function of three variables namely boarding passengers, alighting passengers and passengers on-board using empirical data. In their study they mentioned that there are other components have to be estimated in order to estimate the bus dwell time. In their work they used two techniques; frequentist approach and stochastic approach to model the BA times. Also, the researchers provide a generalized model that is capable to estimate the BA time regardless the bus capacity. Then, they presented different stochastic techniques to estimate the BA times.

### Literature Review for Vehicle Acceleration/Deceleration Modeling

Modelling the acceleration/deceleration has been carried out for the past few decades. Several studies were conducted to predict maximum and typical acceleration levels, considering different vehicle types. These models are mainly categorized into two groups based on the driver's reaction towards the leading vehicle. These groups are car-following models, where the driver's behavior towards the lead vehicle is considered (also known as kinematics models) and dynamics models, where the acceleration/deceleration behavior is not determined by considering

Alhadidi and Rakha 4

the car-following behavior, but by considering the vehicle dynamics characteristics instead. However, the dynamic models are more complex in mathematical representation and development.

Moreover, the current state-of-art and state-of-practice models were developed based on using outdated, limited data or using a traditional measurement method to provide an estimate of vehicle speed ( (4-11), (12), (13), (14)). Yet, none of them focus on transit vehicles based on field data due to the cost of collecting the required parameters from the field and some technical challenges, including data storage and/or data accuracy problems. Recently, the Hydraulics, Electrical and Mechanical (HEM) Data Corporation provided a data logger that collects HEM records from the transit vehicle based on either 2 Hz or 5 Hz frequency. This HEM data provides several parameters of vehicle characteristics and stores these parameters to be used in modeling transit vehicle movement.

*Kinematic Models*

As mentioned earlier, kinematic models focus on the relationship between the moving vehicle and the leader vehicle. In other words, these models consider the relationship between speed, traveled distance, and acceleration/deceleration rates without considering the forces that cause the motion. The simplest model uses a constant acceleration rate along the entire trip; however, this model is unable to express real vehicle movement because vehicle acceleration isn't constant, but it changes over trip time (13). Other researchers argue that vehicles achieve a higher acceleration rate while they are traveling at low speed, compared with the associated acceleration rate at a higher speed. Considering this behavior, a dual-regime model that considers two acceleration rates was proposed; these rates are a constant, higher value at lower speeds and a lower value at higher speeds (10).

Another simple and commonly used model is Drew's model. This model assumes that a vehicle starts at a speed equal to zero with the highest acceleration rate, and this rate decreases linearly as a function of speed (13). However, studying deceleration behavior of different vehicles using field data shows that the observed deceleration rate is lower than the suggested deceleration rate proposed by both the American Association of State Highway and Transportation Officials (AASHTO) and the Institution of Transportation Engineers (ITE). Rather, the observed rate increases as the initial speed increases (11, 13).

Describing the non-uniform behavior of deceleration rate, a polynomial relationship between acceleration and speed was proposed by Akçelik and Biggs (5). This model shows that a vehicle traveling at high speed covers a longer deceleration distance before reaching its target speed.

*Dynamic Models*

Searle's model was the first model that considered vehicle's engine power. However, it overlooks the impact of resistance forces (6). Moreover, the model is valid for estimating the maximum vehicle acceleration rate based on using six vehicles' field data.

The results of Searle's model confirm that both the behavior and trend of the maximum acceleration is similar to the observed field data. However, Rakha and others (4, 8) developed a dynamic model to estimate the maximum vehicle acceleration assuming constant vehicle power.

Another study conducted by other researchers used Global Positioning System (GPS) data from 100 passengers' vehicles to model their deceleration behavior at a controlled stop sign. However, results of this study do not include vehicle characteristics that affect deceleration values (11).

**Literature Review for Re-Entry Delay**

The Transit Capacity and Quality of Service Manual (TCQSM) defines re-entry delay as the time that the transit vehicle spends between door closing and the bus starting to leave the bus stop. This time mainly has two components, namely: the startup time and the clearance time. The startup time is defined as the time needed to begin moving and travel on its path, which can also be called acceleration time. Alternatively, the clearance time includes the re-entry delay (merge-delay), which depends on the traffic volumes in the curbside lane. Typically, merge-delay time increases as traffic volume increases (2). This time depends also on the queued traffic and the upstream traffic signal, which may cause long gaps in traffic or extra delays (15). Transit agencies do not recommend using off-lines bus stops to eliminate the merge-delay time; however, off-line stops reduce rear-end collisions between transit vehicles and the traffic in the adjacent lane. The TCQSM suggests that an average re-entry delay depends on the adjacent lane volume based on 12 buses stopped at un-signalized intersection, where a critical gap of 7 seconds and random vehicle arrival times are assumed (2).

## METHODOLOGY AND DATA COLLECTION

This section handles the different dataset that used in this research work separately. It starts with a description of the collection, reduction, and processing, then it follows by the adopted methodology for the analysis of each dataset.

### Data Description

Here, the collection, reduction, and processing for each dataset is presented below.

#### *Deceleration and Acceleration Times*

The data for modeling acceleration/deceleration times were collected by Blacksburg Transit (BT) by driving buses around town of Blacksburg, VA. The data were collected using a HEM logger that records data with a frequency of 2 Hz. In total, about 47 parameters were collected, 5 of which were used in this study, namely: the time stamp, speed, and GPS location (i.e., longitude, latitude, and altitude). In order to cover a wide range of real-world driving conditions and to determine the uphill and downhill sections, a test was conducted on two road sections: US 460 business highway (with a speed limit of 65 mi/h [104 km/h]) and local streets (with the speed limit from 25 mi/h to 45 mph [40-72 km/h]). As described previously, the HEM logger records data with a frequency of 2 Hz. In order to generate a second-by-second record, the original data were combined by averaging the data points within each individual second. After generating second-by-second results, the null data where the longitude/latitude/altitude had a value of "0" were removed. Then, an approach was developed to extract the vehicle trajectories.

The extraction logic for managing the HEM data was built in-house. The input data was the HEM data with the second-by-second time stamp, GPS parameters, and bus speed. The logic tracks second-by-second data and checks the trend in the recorded speed value. The speed trend changes were then split into either acceleration or deceleration, depending on the trend.

A program was used that checks the data compatibility with two-stage validation standards. These standards are either at the point level or at the entire trip level. Trip or point level data that failed to meet these standards were purged. At the point level, observations are deleted for the following reasons: (a) recorded "0" from the GPS data and/or (b) data points collected at garages or check points. After omitting the data points, the software split each group of points into either acceleration or deceleration events based on the speed trend. The validation steps at the trip level followed these steps: (a) split the trip into acceleration or deceleration based on the speed trend (i.e., if the speed trend is ascending that means the trip is acceleration trip and if the speed trend is descending that means the trip is deceleration trip) and (b) all the trips used

Alhadidi and Rakha                                                                                                                6Alhadidi and Rakha                                                                                                               6 in this work were either ascending from or descending to zero for either the acceleration or deceleration trips, respectively.

*Re-entry Delay time*
In order to estimate re-entry delay, different microscopic simulation scenarios were created using INTEGRATION. Different scenarios were run by varying the volume from 600 veh/hr up to 2,500 veh/hr at 50 veh/hr increments, speed variability changes from 0 to 0.015 at 0.025 increments, a fixed capacity equal to 2,500 veh/hr/ln, and the free-flow speed varying from 35 km/hr to 55 km/hr at 5 km/hr increments. A total of 1,365 simulation runs were performed in this study in order to capture different values of the re-entry delay time under different traffic conditions. Figure 3 shows the geometric configuration of the tested network. The network consists of three connected links, all of them with the same capacity and same traffic density at capacity. Link 2 has the bus stop and only allows the transit vehicle to use it. Thus, the bus is held for a pre-defined fixed stopping time to serve the demand at the bus stop, then the bus waits for a certain time to merge with the traffic in the adjacent lane to finish its trip.

      The generated output file that tracks each vehicle traveling along each link second-by-second, checking for extra stopped time (i.e., the time that a transit vehicle spends beyond the pre-defined time) and the traffic passing the stopped bus until the end of link 2. In other words, the tracking was done based on the flow and the density in the bus stop's adjacent lane.

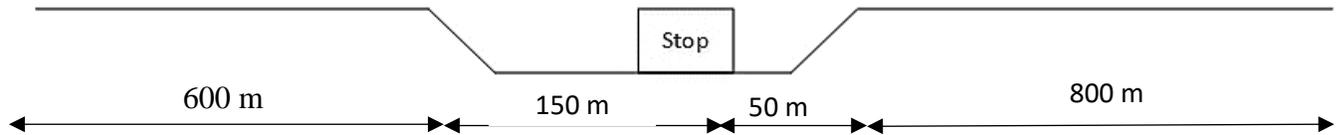

**Figure 3: Network Configuration**

**Methodology**
This section presents the research methodology. As was mentioned in the literature review, the time elapsed between the bus stopping and the beginning of passenger boarding, or alighting is beyond the scope of this work. Instead, the methodology described here handles the deceleration time, re-entry delay time, and the acceleration time.

*Deceleration Time*
Following the trend split, the deceleration trips were extracted. The extraction logic summarizes it via four parameters: trip time, deceleration rate, initial speed, and traveled distance.

      Using vehicle kinematic equations, the deceleration rate was computed using Equation(1). Then, the computed deceleration rate was used to estimate the deceleration rate as a function of the initial speed, as the final speed equals zero. Thereafter, to compute the deceleration time, the deceleration distance should be calculated first. The predicted deceleration rate, which is presented in Equation (2), was applied to the well-known deceleration distance equation in order to compute the deceleration distance, which is presented in Equation (3).

$$Deceleration\ Rate = \frac{V_0}{t} \qquad (1)$$

$$a = \frac{B_1}{B_2 + B_3 V_0 + V_0^{-B_4}} \qquad (2)$$

$$Deceleration\ Distance = \frac{V_0^2}{2 * g(a \mp G)} \qquad (3)$$

Here, $V_0$ is the initial speed which is measured in (km/hr), $t$ is the entire deceleration maneuver time, $a$ is the deceleration level, $g$ is the standard acceleration due to gravity (which equals to 9.81 m/sec²), and $G$ is road grade, which is computed using Equation (4).

$$G = \frac{elevation(t) - elevation(t - \Delta t)}{\sqrt{\big(D(t) - D(t - \Delta t)\big)^2 - \big(elevation(t) - elevation(t - \Delta t)\big)^2}} \qquad (4)$$

Where $elevation(t)$ is the altitude (m) at time (t) and $(t - \Delta t)$; and $D(t)$ is the distance (m) a bus traveled at the interval of $\Delta t$. The elevation was collected by HEM and the traveled distance was computed based on the collected speed data using Equation (5).

$$Distance = V_0 \times \Delta t \qquad (5)$$

*Re-entry Delay Modelling*
In order to model the re-entry delay time, several scenarios were generated using INTEGRATION as a microscopic simulation software. INTEGRATION produces a second-by-second probe vehicle output file, where each vehicle movement is tracked second by second on each link. At each generated data point, much information can be extracted from these data reports, such as the time stamp, vehicle location, speed, distance, delay, total delay, and vehicle emissions. In this study, the base scenario and the generated scenarios were ran and the output data were extracted. In the base scenario, the network was run while it has only the transit vehicle. While, the generated scenarios were run separately according to the variations that were mentioned before. For each scenario, the transit travel time from leaving the bus stop until it reaches the end of the link in addition to the density on the adjacent lane were extracted. In this study, two statistical models were developed, one using classical frequentist modeling and another using stochastic modeling based on a Bayesian approach. In both of the two developed models, the re-entry delay was considered as the dependent variable since density is likely to influence the re-entry delay time.

**Re-Entry Delay Modelling Using Classic Frequentist Statistics.**
Modeling the re-entry delay using classic frequentist statistics (CFS) was done using the generated data from INTEGRATION using the generalized linear regression (GLR) for modeling the linear relationship between the dependent variable and independent variable(s). GLR assumes the linear relationship between model variables and the uses least squares error estimation (LSE) to estimate the model's coefficient. The general form of the GLR is presented in Equation (6)

$$y = \beta'X + \epsilon \qquad (6)$$

Where $y$ is the dependent variable, $X$ is a vector of explanatory variable(s), $\beta$ is the vector of estimable parameters or coefficients, and $\epsilon$ is the random error term.

In LSE, the model parameter coefficients are estimated based on minimizing the sum of the squared difference between the observed value and the predicted value. Consequently, the GLR assumptions should be checked. Model assumptions can be checked using either graphical

Alhadidi and Rakha                                                                                   8or appropriate diagnostic tools, including a homoscedasticity check and check for a normally distributed error term.

**Re-entry Delay Modelling Using Bayesian Approach**
Using classical frequentist statistics (CFS) assumes that the model has a constant variance, such that any information about the model's variance is determined during model building. Moreover, the variables in empirical studies, according to asymptotic theory, has a stochastic nature. This stochastic nature arises from the fact that the explanatory variable in some studies becomes a response variable in another study. Thus, CFS is unable to detect any change in the different statistical inferences.

Another key important fact to using Bayesian modeling for the re-entry delay parameters is that the Bayesian approach is able to use a prior knowledge for a certain case to be expanded to other different conditions including; different bus characteristics, different network configurations, and different prevailing traffic conditions. If this is the case, then the MLR model coefficients are no longer accurate. Instead, we can use the MLR model coefficients as the prior coefficient estimator. In this case, the agency needs to collect limited local data, the model would then use the local data together with the prior coefficients to estimate a posterior set of coefficients that are calibrated to the locality. In other words, the Bayesian approach can modify the model parameters to reflect the local conditions.

Another reason to use the Bayesian approach for other localities is that the Bayesian method provides a more direct expression of uncertainty, including complete ignorance. Also, in the Bayesian approach, there is no worry of using a smaller sample size, as it uses a smaller number of parameters. More importantly, the Bayesian method has the capability of handling non-normal parameters, as Bayesian methods provide more accurate results because they can deal with asymmetric distributions.

Generally, the maximum likelihood estimation (MLE) is used in the classical frequentist inferences. MLE assumes model parameters that are fixed and known, and the bottom line of the selection of these parameters emerges from maximizing the likelihood function for the selected model. Notably, using informative knowledge about the process for both prior and posterior information about the process, using Bayes' theorem (16) in Equation (7). This equation can be expanded to achieve stochasticity among model parameters, which uses the conditional probability formula in Equation (8).

$$posterior\ \alpha\ prior\ \times likelihood \quad (7)$$

$$p\ (y\mid m) = \frac{P\ (\beta,\ \sigma\mid m)\ P\ (y\mid X, \beta,\ \sigma,\ m)}{P\ (\beta,\ \sigma\mid y,\ X, m)} \quad (8)$$

Where:
$p\ (y\mid m)$ = the marginal distribution of y.
$P\ (\beta,\ \sigma\mid m)$ = the prior model distribution.
$P\ (y\mid X, \beta,\ \sigma,\ m)$ = the likelihood estimation.
$P\ (\beta,\ \sigma\mid y,\ X, m)$ = posterior distribution.

The selected model parameters can be calibrated using the Bayesian statistics method. Associated with using the Bayesian method of data analysis is the Monte Carlo approach, which has been used to generate the posterior distributions when these distributions cannot be generated analytically. The Bayesian parameters can be modeled using the Markov Chain Monte Carlo (MCMC) slice algorithm implemented with the Bayesian algorithm in the R software. This algorithm is designated for use in the distribution of an arbitrary density function to sample the data using a known constant of proportionality, which is needed to sample from a complicated posterior distribution whose normalization constant is unknown. The approach generates random samples from these distributions to estimate the posterior distribution or derived statistics (e.g., mean, median, and standard deviation).

The MCMC slice algorithm requires an initial solution for the model parameters, the prior distribution, and the residual variance. With these priors, the posterior distribution of model parameters conditionally depends on the response variable and the variance. Transit agencies can use this information to draw the marginal distribution using the Monte Carlo approach by first, drawing the variance from its marginal distribution.

**Boarding and Alighting Times**
As it was mentioned earlier, the researchers estimated the BA times in an earlier study using empirical data from king county (3) . In their work two multilinear models were developed to estimate BA times regardless the bus capacity namely; frequentist method and stochastic method.
Speaking of estimation of passengers' BA times, they found that it depends on three variables namely; the number of alighting passengers, the number of boarding passengers, and the number of passengers on board the bus.  The models show that passenger BA times increase with an increase in all three variables (i.e. positive model coefficients). Also, they make sure that the estimation of the BA times doesn't violate the TCQSM limits by providing lower and upper limit for the BA estimation formula which is presented in Equation (9)

$$T_{BA} = \begin{cases} Min_{Stop} & X_1 + X_3 = 0 \\ Min\left(e^{\beta_1\sqrt{X_1}+\beta_2\sqrt{X_2}+\beta_3 X_3}, Max_{Stop}\right) & X_1 + X_3 > 0 \end{cases} \quad (9)$$

Here $T_{BA}$ is the duration of time the transit vehicle door is open (BA time), $\beta$i's are the model coefficients, $X_1$ Is the of the ratio of the number of boarding passengers to the bus capacity, $X_2$ Is the ratio of the number of passengers on board the bus to the bus capacity, $X_3$ is the ratio of the alighting passengers to the bus capacity (3).

**RESULTS**
This section presents the research results. The results are categorized into three main groups: deceleration profile, acceleration time, boarding alighting times, and clearance time.

**Deceleration Profile**
This section presents trip deceleration characteristics including deceleration rate, deceleration distance, and deceleration time.

*Deceleration Rate*
The idea of conducting this research arose from the fact that the power delivered by the engine during braking cannot be assumed to be the maximum power nor a constant percentage of the power along the deceleration trip. Moreover, a transit vehicle driver does not typically impose a high brake pedal input that fully stops the transit vehicle; rather, the driver must impose a gradual deceleration rate in order to minimize the vehicle jerk (rate of change of acceleration). Subsequently, this encouraged the research team to adopt vehicle kinematics equations, as they assume a constant acceleration/deceleration rate that estimate the deceleration rate as a function of the speed change over the entire trip; in other words, the research team agreed that the transit vehicle maintains a constant deceleration rate during stopping maneuvers so that passengers' stability is not affected by the vehicle. However, modeling the deceleration rate as a function of speed is a challenging task because, once the driver reaches their comfortable deceleration rate, the deceleration rate increases as the speed decreases to assure that the vehicle is completely stopped before passing the desired deceleration distance, and typically this rate is higher than the deceleration rate at a higher speed. In fact, the driver may start to decelerate from a certain speed by gradually pushing on the brake pedal until they reach a speed level at which they are assured the vehicle will stop; otherwise, they may only have to barely push on their brake pedal to make sure that the vehicle stops within the target distance.

The research team investigated several functions to model the deceleration rate as a function of the initial speed, since the quadratic model is unable to handle the model characteristics around the extreme points, and the linear regression model cannot produce a reasonable prediction (as the estimated value does not provide a good measure-of-fit). After a thorough investigation, the function in Equation (10) was found to be the best fit to this research data. The proposed model shows its capability of handling the extreme points and has a degree of freedom around the extremes.

$$Deceleration\ Rate = \frac{2.7 * e^{-5}}{-2.6e^{-8} + 2.618e^{-6}V_0 + V_0^{-22.18}} \quad (10)$$

Model validity and adequacy were checked using different means, including graphical tools and appropriate diagnostic, as ways to check the selected model using the normal quantile plot and check the homoscedasticity. The chosen model has to show approximate residual errors in additional to having a random error pattern, as shown in Figure 4.

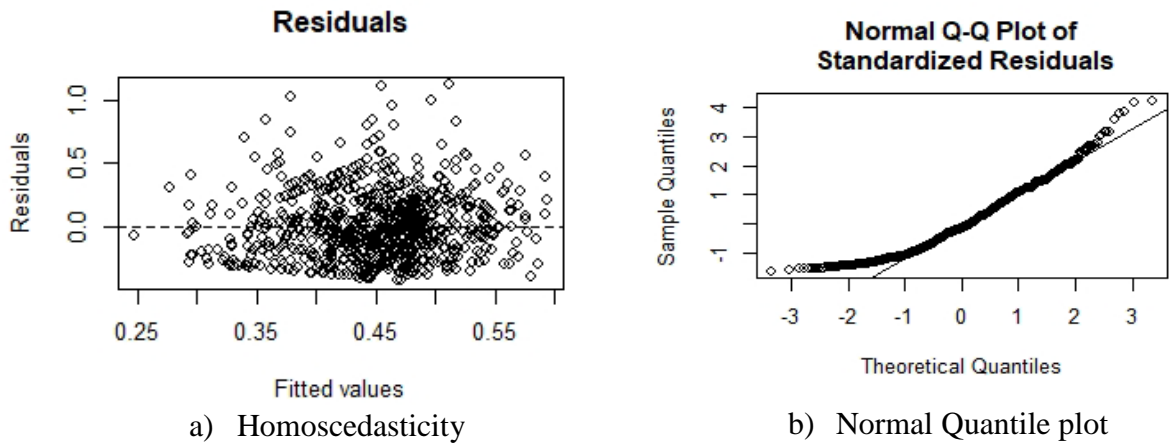

a) Homoscedasticity      b) Normal Quantile plot

**Figure 4 : Model Diagnostic**

Figure 4.a shows the model's homoscedasticity check. It can be seen that the residual errors are independent from each other and are randomly distributed around 0. It also shows that these values do not exceed 1. Figure 4.b shows that the error term approximately follows normal distribution. However, it also shows that model results in large values at the tails of the distribution compared with standardized model assumptions. This issue was resolved by setting the minimum deceleration rate as 0, which corresponds with the value of 0 for speed change. In our case, while there was no change in speed, we assumed that the bus maintained a fixed deceleration rate and thus, set the deceleration value of zero.

*Deceleration Distance*
Deceleration distance was computed based on Equation(3) using the estimated deceleration rate in Equation (10). The computed distance was compared against the field distance, and both the mean absolute percentage error (MAPE) and root-mean-square-error (RMSE) were computed and found to be 2.2% and 28.36%, respectively.

*Deceleration Time*
The deceleration time was computed using the computed deceleration distance according to Equation (3) and using the kinematic equation, which is shown in Equation (11).

$$Deceleration\ Time = \frac{Deceleration\ Distance}{V_0} \quad (11)$$

Comparing the actual deceleration time with the estimated time was done using a graphical tool and then by computing MAPE and RMSE. The MAPE and RMSE were found to be 0.3% and 13.3%, respectively.

*Boarding alighting times*

The coefficient of the BA estimation model that was presented in Equation (9) along with their *p* values was presented in the previous study (3).

Transit agencies wanting to adapt the model may be categorized as one of two scenarios. First, the transit agency has the same conditions of King County. In this case, the developed model in Equation (9)with its coefficients in (3) can be used directly. The second scenario is where the transit agencies have different local conditions (i.e. different number of bus doors) or different passenger characteristics (i.e. cyclists, strollers, elderly people, or passengers with disabilities), and different fare payment methods. If this is the case, the model coefficients estimated in Table 1may no longer be accurate. However, the model will still be applicable if a Bayesian approach is implemented to update the model coefficients  (the generalized one), as the prior(3). The agency would need to collect some limited local data, the model would then use the local data together with the prior coefficients to estimate the posterior of coefficients that are calibrated to the locality. In other words, the Bayesian approach can combine information from the existing model fitting results and locally collected data to generate posterior distribution of model parameters that reflect local conditions.

To be more specific, the posterior of the developed BA model can be used as a prior to develop locally specific model coefficients. This is based on the assumption that the multi-parameter posterior from previous studies can be accurately represented by independent marginal distributions of each parameter. Thus, a transit agency can use posterior of a previous study as the priors to reflect common characteristics of transit.  When combined with the locally collected data, the output posterior reflects both common characteristics from existing studies and local specific information.

**Clearance Time**
This section presents the transit vehicle waiting time to merge with the traffic. As was mentioned before, this time has two components: the startup travel time and the re-entry delay time. This section discusses the re-entry delay, while the startup travel time is presented in the acceleration time.

*Re-Entry Delay*
In this section, modeling the re-entry delay using classic frequentist statistics and the Bayesian approach are presented.
**Modelling Re-entry Delay using Classic Frequentist Model**
The re-entry delay time measures the extra time that a transit vehicle has to wait at the bus stop to merge with the traffic in the adjacent lane. This time is expressed as a function of the traffic density in the adjacent lane. Estimating the re-entry delay time as a function of density was done using generalized linear modeling (GLM). Assuming that the merge delay is a continuous variable, and the traffic density is an integer variable. The general linear model was fit to the data, considering a normal distribution of the results from Equation(12) .

$$Re - entry\ Delay = B_0 + B_1\ K \qquad (12)$$

Where $K$ is the density (Veh\km\ln); and B's are the different model coefficients. Using LSE, the model coefficient values along with their p values are summarized in Table 1.

**Table 1: Statistical Interference for Merge Delay Model**

| Parameter | Estimate | Standard Error | P-value |
|---|---|---|---|
| $B_1$ | 1.788 | 0.050 | <0.001 |
| $B_2$ | 0.318 | 0.004 | <0.001 |

Table 2 shows the re-entry model parameters' statistical interferences. The transit vehicle has to wait at least 1.788 seconds to re-enter traffic. This time is close enough to the average perception time, which is 1.5 sec (17). However, this time increases by 3.1 seconds per 10 vehicle/km/ln increase in density in the adjacent lane.

The model capability to explain the variance in the data was checked using goodness-of-fit was 92%. Moreover, the adequacy and validity of the model coefficient estimation was done using the general linear model assumptions. Choosing the best model among a set of models means that the selected model should have a normal residual error, as shown in Figure 5.

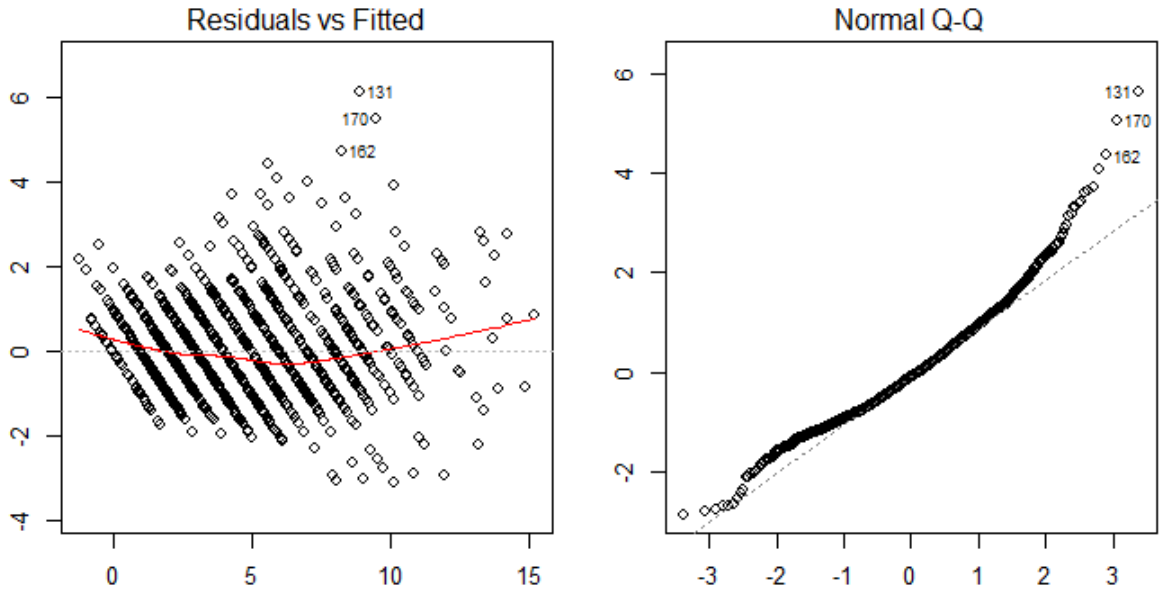

**Figure 5: Re-entry Delay Model Diagnostic**

Figure 5 shows the re-entry delay time model diagnostic for both the residual error and the normal quantile plot. Both the residual plot and normal quantile plot show that the residual errors are normally distributed around zero and follow normal distribution.

*Acceleration Time*

As a reminder, the acceleration rate was computed for the entire trip in addition to the second-by-second acceleration rate using the kinematic equation, which was presented in Equation (1). Then, two acceleration rate values were extracted: the average acceleration rate and the maximum acceleration rate for the entire trip. The reason behind selecting these two values is that, during the entire trip, the acceleration rate does not stay constant, nor does the engine reach its maximum power. In fact, different engine powers with different driving conditions result in different maximum acceleration rates to reach the maximum desired speed. According to the dataset used, the maximum speed was 70.4 km/hr. In order to generalize the prediction model, the acceleration rate had to be transferrable to any transit vehicle and any driving condition. The acceleration rate prediction model is presented in Equation (13).

$$\frac{a_{average}(U)}{a_{max}} = \left(1 - \left(\frac{x}{2.554}\right)^{2.519}\right) \qquad (13)$$

Where $a_{average}(U)$ is the average acceleration rate that can be obtained during the acceleration event; $x$ is the ratio between the target speed and the maximum speed =(V/70.4); and $a_{max}$ is the maximum acceleration rate that can be obtained during the acceleration event.

The plot of the model of Equation (13) is shown in Figure 6. The proposed model shows its capability to capture the variation in the data with a goodness-of-fit equal to 0.84.

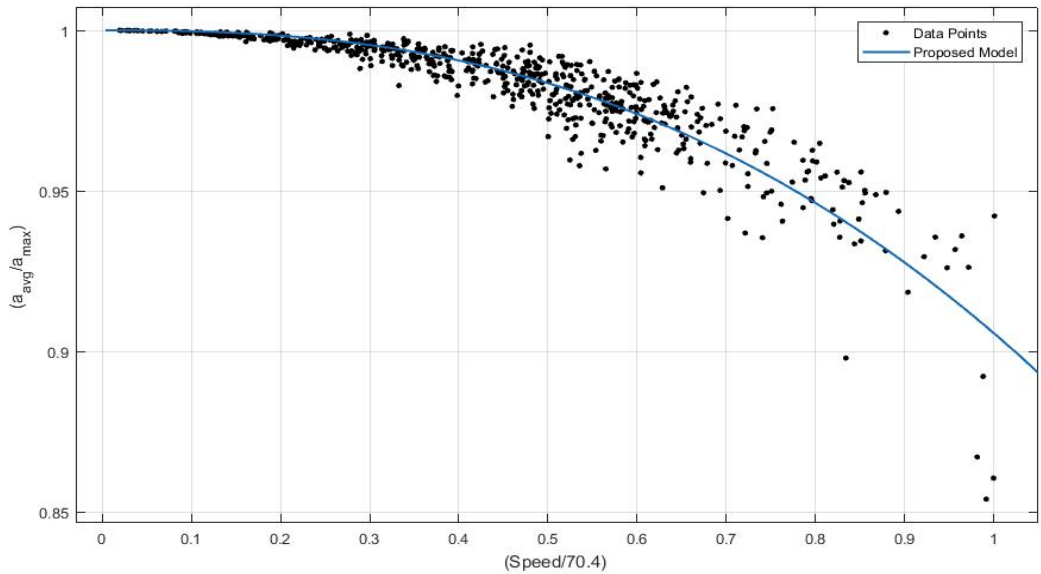

**Figure 6: Acceleration rate prediction model.**

In order to estimate the acceleration time, the predicted acceleration rate in Equation (13) can be used in Equation (14).

$$Acceleration\ Time = \frac{V_{target}}{a_{avg\_predicted}} \quad (14)$$

Where the $V_{target}$ is the desired speed to which the transit vehicle wants to accelerate. Comparing the estimated acceleration time against the field value, both RMSE and MAPE were computed and found to be 2.72% and 0.42%, respectively.

**Modelling Re-entry Delay Using Bayesian Approach**
The selected model parameters can be calibrated using the Bayesian statistics method. Associated with using the Bayesian data analysis, the Monte Carlo approach has been used to generate the posterior distributions when these distributions cannot be generated analytically. More specifically, the Bayesian parameters can be modeled using the Markov Chain Monte Carlo (MCMC) slice algorithm implemented with the Bayesian algorithm in R software. This algorithm is designed to use a distribution of an arbitrary density function to sample the data using a known constant of proportionality, which is needed to sample from a complicated posterior distribution whose normalization constant is unknown. The approach generates random samples from these distributions to estimate the posterior distribution or derived statistics (e.g., mean, median, and standard deviation).

The MCMC slice algorithm requires an initial solution for the model parameters. With these priors, the posterior distribution of the model parameters is conditionally dependent on the response variable and the variance. Then, the researchers used this information to draw the marginal distribution using the Monte Carlo approach by first drawing the variance from its marginal distribution. Notably, the posterior distribution of the variance distribution is scaled through an inverse Chi-square distribution with scaled deviation and (n-k) degrees of freedom, where n is the number of observations and k is the number of model parameters. In this case, the number of observations is 1,365, and (1,365-2=1,363). Table 2 summarizes the different statistics for the 1,000 realizations for the model parameters. It can be seen that the mean value of all the model parameters are close to the parameter estimates from the classical frequentist GLM model in Table 1.

**Table 2 : Summary Statistics of the Different Model Parameters**

| Parameter | Mean ($\mu$) | Quantiles | | St. Dev. ($\sigma$) | Skewness ($\gamma_3$) | Kurtosis ($\gamma_4$) |
|---|---|---|---|---|---|---|
| | | $Q_{0.025}$ | $Q_{0.975}$ | | | |
| $B_0$ | 1.789 | 1.604 | 1.823 | 0.050 | -0.017 | -0.030 |
| $B_1$ | 0.3185 | 0.303 | 0.321 | 0.004 | 0.008 | -0.016 |

Moreover, measuring the randomness between the generated samples as a function of time was done using autocorrelation functions (ACF) for modeling the coefficients, which is presented in Figure 7. As can be seen in the figure, the solid horizontal lines represent the 95% confidence limits, while the stem lines represent the autocorrelation function. As would be expected, an autocorrelation of 1.0 is observed for a lag of 0, while the values are within the confidence limits for larger lag times.

In addition, a Kolmogorov-Smirnov goodness of fit test (K-S test) was run on the model parameter distributions and concluded that there was insufficient evidence to reject the null hypothesis, as the model parameter distributions were different from the normal distribution at a level of significance of 0.05, as illustrated in Figure 8 .

Another important illustration of the model results is presented in Figure 9, which shows the interaction between the model parameters ($\beta_0$, $\beta_1$). When comparing $\beta_0$ to $\beta_1$, it is clear that a linear relationship between $\beta_1$ and $\beta_0$ can be deduced.

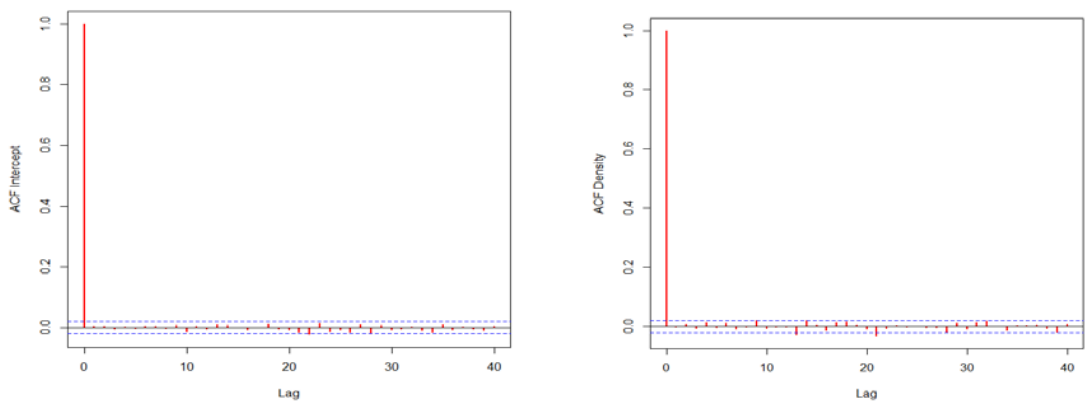

**Figure 7 : Variations in the Autocorrelation Function of the Model Parameters**

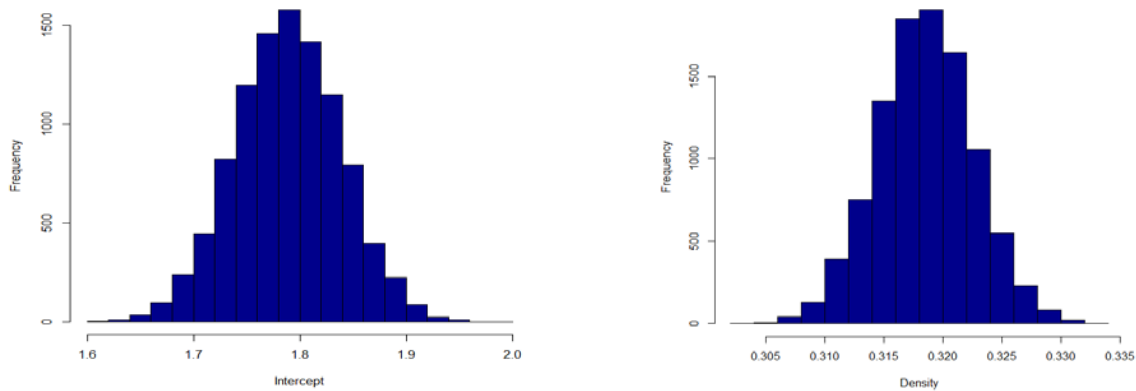

**Figure 8: Histograms of the Model Parameter Distributions**

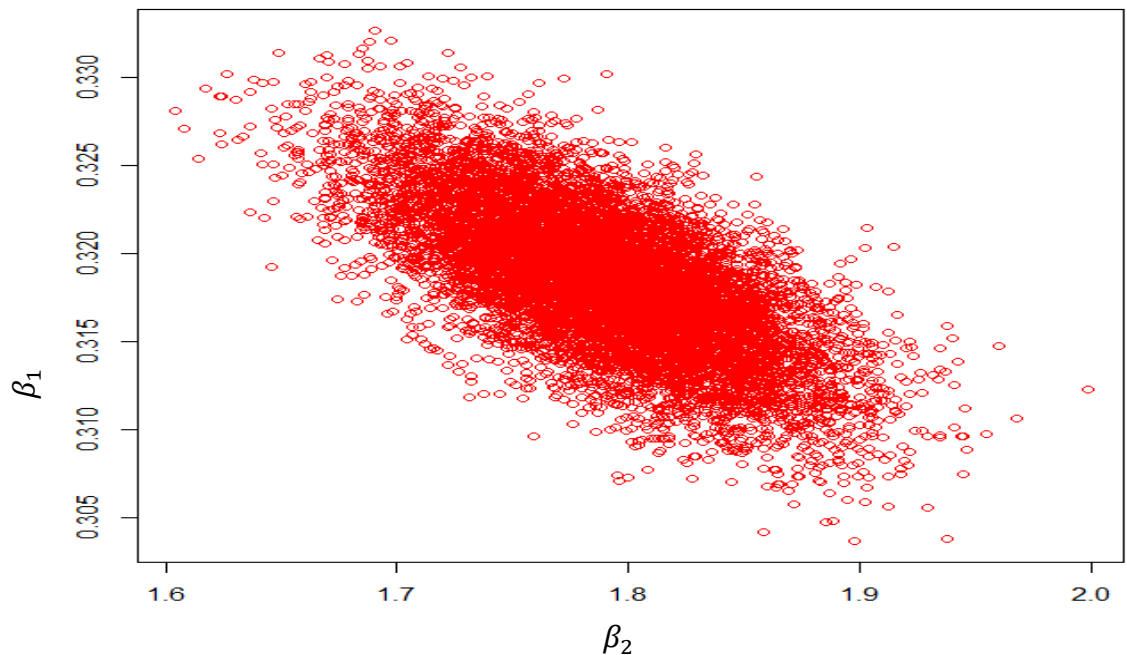

**Figure 9: Inter-dependence of the Bayesian Model Parameters**

## DISCUSSION
This work primarily focuses on the deceleration time and the clearance time (as the loading time has been studied before by the researchers). Thus, this work quantifies the total delay time components at bus stops. The total delay has three components: deceleration time, loading time, and the clearance time. The research team adopted vehicle kinematic equations to estimate the deceleration time. Also, the researchers computed the clearance time as a function of the link density and the targeted speed.

Alhadidi and Rakha															17The results of this work show the validity of using the adopted methodology as it compared the estimated deceleration distance and time with the field values. Also, the results indicate that the re-entry delay variable confirms the hypothesis variable signs.

## CONCLUSIONS AND RECOMMENDATIONS

Modeling transit dwell time has been done over years ; however, due to congestion, the dwell time does not reflect the full boarding/alighting passenger time. Instead, it has other components including the time required to completely stop the transit vehicle and the time required to merge with the traffic again.

To test the study hypothesis, the researchers started by adopting a kinematic equation to express the transit vehicle deceleration and acceleration characteristics due to the fact that the transit vehicle cannot maintain a constant acceleration/deceleration rate or a maximum acceleration/deceleration rate during for the entire maneuver. Hence, developing a mathematical model that is able to predict the acceleration/deceleration rate comes as the first priority. Consequently, the deceleration distance can be computed using the predicated deceleration rate, after which the deceleration time can be estimated more easily. That being said, the adopted methodology was compared against empirical data, and both the estimated deceleration distance and the deceleration time show their ability to replicate the field data with a reasonable error. The acceleration rate was computed as a function of vehicle and driver characteristics and, by revisiting the predicted acceleration model, the roadway characteristics are captured by the free-flow speed. Conversely, vehicle characteristics are measured by the maximum acceleration rate. Using the predicted acceleration rate and the change in speed, the acceleration time was computed using the kinematic equations.

Evaluating the adopted methodology for computing both the acceleration and deceleration time was done using the MAPE and RMSE. Overall, the study results show the validity of the adopted methodology to estimate the deceleration time; by comparing the estimated deceleration time with the field data, the MAPE and RMSE were computed to be 0.3% and 13.3%, respectively.

Also, the re-entry delay was estimated as a function of the traffic stream density in the vicinity of the bus stop and the change in transit vehicle speed, such that the total lost time at a bus stop includes the boarding/alighting time, in addition to the re-entry delay and the time required to decelerate and accelerate to its desired speed.

It is worth mentioning that the study results show that the vehicle kinematic equations help in estimating the deceleration time and acceleration time respectively, but also that the factors that affect the clearance time have a reasonable sign. The results of this study can be used to update the transit vehicle arrival and departure times at bus stops and provide a better estimation of bus arrival times at traffic signal. Also, it can be used to update traffic signal and transit stop configurations.

Although the study meets its objective, the study has some limitations. For example, the effect of pavement and weather conditions on the vehicle kinematics were not considered in this study. This is because the main objective was to estimate the deceleration time alone. However, these factors may affect the deceleration rate.

## ACKNOWLEDGEMENTS
This research was funded by the US Department of Transportation through the University Mobility and Equity Center (Award 69A3551747123). Alhadidi is a Jordanian fellow supported by Al-Ahliyya Amman University (AAU).

## AUTHOR CONTRIBUTION STATEMENT
The authors confirm contribution to the paper as follows: study conception and design: Alhadidi, Rakha; data reduction: Alhadidi; analysis and interpretation of results: Alhadidi, Rakha; draft manuscript preparation: Alhadidi, Rakha. All authors reviewed the results and approved the final version of the manuscript.